# The quantitative and qualitative content analysis of marketing literature for innovative information systems: the Aldrich Archive

## Authors: Sebastian Fass[1] and Kevin Turner[2]

**ABSTRACT**

The Aldrich Archive is a collection of technical and marketing material covering the period from 1977 to 2000; the physical documents are in the process of being digitised and made available on the internet. The Aldrich Archive includes contemporaneous case studies of end-user computer systems that were used for marketing purposes. This paper analyses these case studies of innovative information systems 1980 - 1990 using a quantitative and qualitative content analysis.

The major aim of this research paper is to find out how innovative information systems were marketed in the decade from 1980 to 1990. The paper uses a double-step content analysis and does not focus on one method of content analysis only. The reason for choosing this approach is to combine the advantages of both quantitative and qualitative content analysis.

The results of the quantitative content analysis indicated that the focus of the marketing material would be on information management / information supply. But the qualitative analysis revealed that the focus is on monetary advantages. The strong focus on monetary advantages of information technology seems typical for the 1980s and 1990s. In 1987, Robert Solow stated you can see the computer age everywhere but in the productivity statistics. This paradox caused a lot of discussion: since the introduction of the IT productivity paradox the business value of information technology has been the topic of many debates by practitioners as well as by academics.

**INTRODUCTION**

Michael Aldrich is one of England's most reputable IT entrepreneurs. For more than four decades he worked in the IT industry and was CEO of ROCC Computers (also known as Redifon and Rediffusion). After his retirement as CEO in 2000 Aldrich became non-executive chairman. He also advised the UK government (in the Thatcher era) on IT-related issues and was an early advocate of the "cabling" of Britain to provide nationwide broadband access.

The Aldrich Archive is a collection of technical and marketing material covering the period from 1977 to 2000; the physical documents are in the process of being digitised and made available on the internet. The Aldrich Archive includes contemporaneous case studies of end-user computer systems that were used for marketing purposes. This paper analyses these case studies of innovative information systems 1980 - 1990 using a quantitative and qualitative content analysis. (For those not familiar with ROCC the degree of innovation may be a surprise – these cases include early examples of home shopping, access to online real-time travel information from home and business-to-business stock location systems).

The major aim of this research paper is to find out how innovative information systems were marketed in the decade from 1980 to 1990. For example, which monetary and non-monetary advantages were highlighted, which keywords were used and are there any patterns within the different case studies?

The paper uses a double-step content analysis and does not focus on one method of content analysis only. The reason for choosing this approach is to combine the advantages of both quantitative and qualitative content analysis (Neuendorf 2002, Krippendorff 2004). This paper does not contain a literature review on innovative information systems from 1980 to 1990; it focuses only on the content analysis of marketing material.

[1] Sebastian Fass works at ERGO Insurance Group, Germany. He formerly worked as Lecturer at University of Brighton Business School, United Kingdom.
[2] Kevin Turner works as Principal Lecturer at University of Brighton Business School, United Kingdom.

Quantitative content analysis is able to deal with a large amount of data (in this case: words); it helps to discover patterns within the data and delivers rather broad research results. The number of occurrences of words and phrases can be an indicator for the emphasis of the marketing material, but it can also be misleading (Stemler 2001). That is where the qualitative content analysis comes into play, because it focuses on smaller amounts of data and delivers in-depth research results (Mayring 2000).

The figure below displays the procedure of the research steps. As shown, the qualitative analysis is an iterative process.

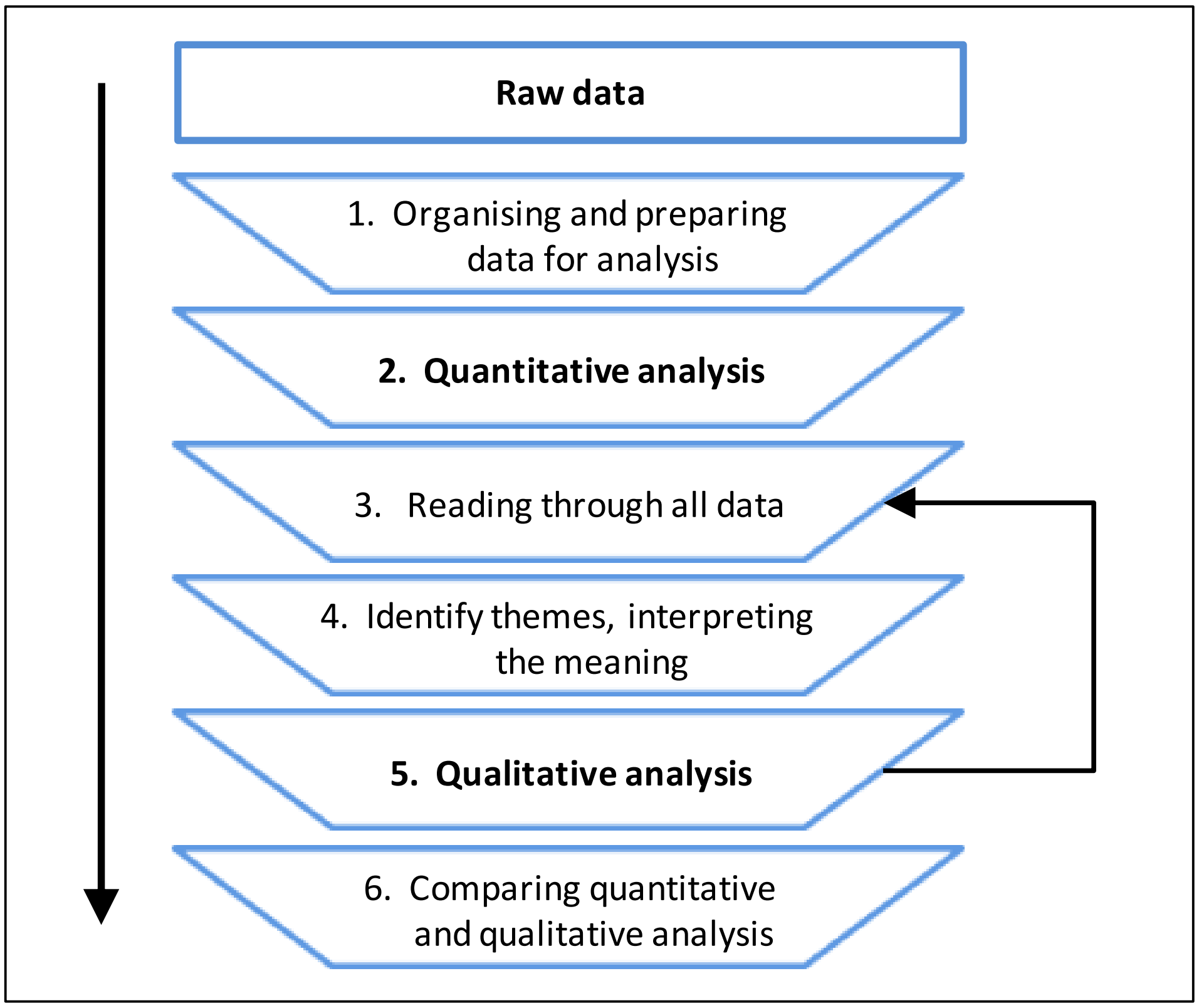

Figure 1: The procedure of the research project

The paper consists of three different sections. The first section deals with the quantitative content analysis and its findings. The second section, the qualitative content analysis and its findings, builds upon the first section. Hence it confirms or contradicts some of the findings of the quantitative content analysis. The third section draws together the findings of both research steps; to reach conclusions it highlights the limitations of the research project and provides recommendations for future research endeavours.

It would go beyond this research paper to discuss general advantages and disadvantages of quantitative and qualitative research methods. The paper attempts to minimise the disadvantages of both research methodologies by combining them. All performed research steps are described in detail to allow the reader to "experience" the procedures of the different research steps (Sandelowski & Barroso 2002).

## QUANTITATIVE CONTENT ANALYSIS

### The procedure of the quantitative analysis

The quantitative analysis started with 29 case studies which were published between 1980 and 1990. All case studies are available in the Aldrich Archive online and can be accessed via the internet at: http://www.aldricharchive.com/innovative_information.html.

First of all the case studies had to be downloaded to the local computer. It needs to be mentioned that the case studies, which were written in the 1980s and early 1990s, were originally available only as hardcopies. In order to make them available online, the hardcopies were scanned and then converted to PDF (Portable Document Format) files. The quality of these PDFs is acceptable to be read by people, but the quality was not good enough to run an automated analysis.

For this reason the case studies had to be transformed into a machine readable format. The programme ABBYY FineReader 9.0 Professional was used to convert the PDF files into one Microsoft Word document. Once again the quality of the PDF was problematic and the Word document had to be corrected by hand. This data preparation phase was time consuming and it took several days to double check the Word document and the case studies. The final Word document has 35,412 words; equivalent to 74 pages.

After the Word document had been prepared for analysis, the programme Primitive Word Counter was used to count the amount of words within the document. The results of this word counter have been visualised with Wordle. It needs to be mentioned that the tag clouds are not able to display the whole 1-word count (for the sake of completeness, the word count is included in the appendix of this paper.).

**Results**

The following tag clouds display the results of the word frequency count. Articles, prepositions, postpositions and other irrelevant words have been removed, because they did not contribute to the aim of the research project (Stemler 2001).

There are three tag clouds in total. After each analysis the most significant word has been removed. The reason for this removal is that the high occurrence of certain words "squeezes out" other words. By removing the most significant word, the remaining words get automatically rearranged.

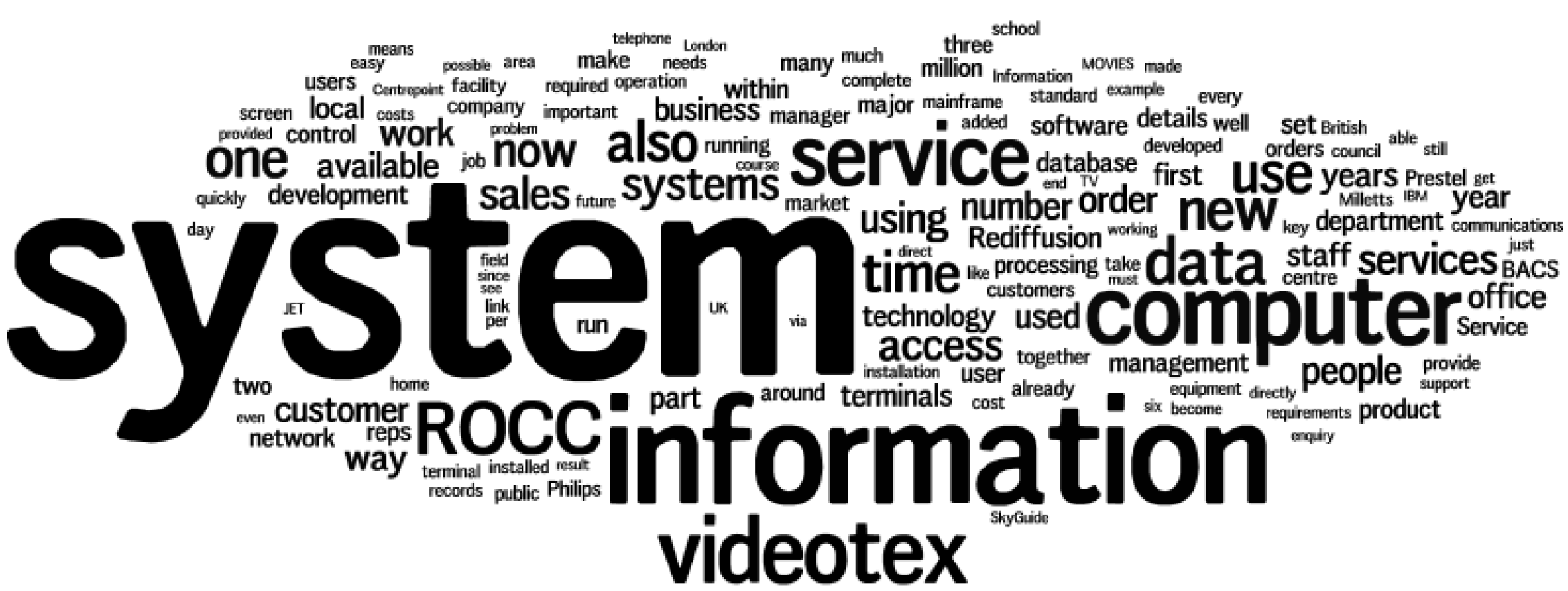

Figure 2: Tag cloud 1 - original version

The tag cloud above displays the original version. In the tag cloud below the words 'system' and 'systems' have been removed.

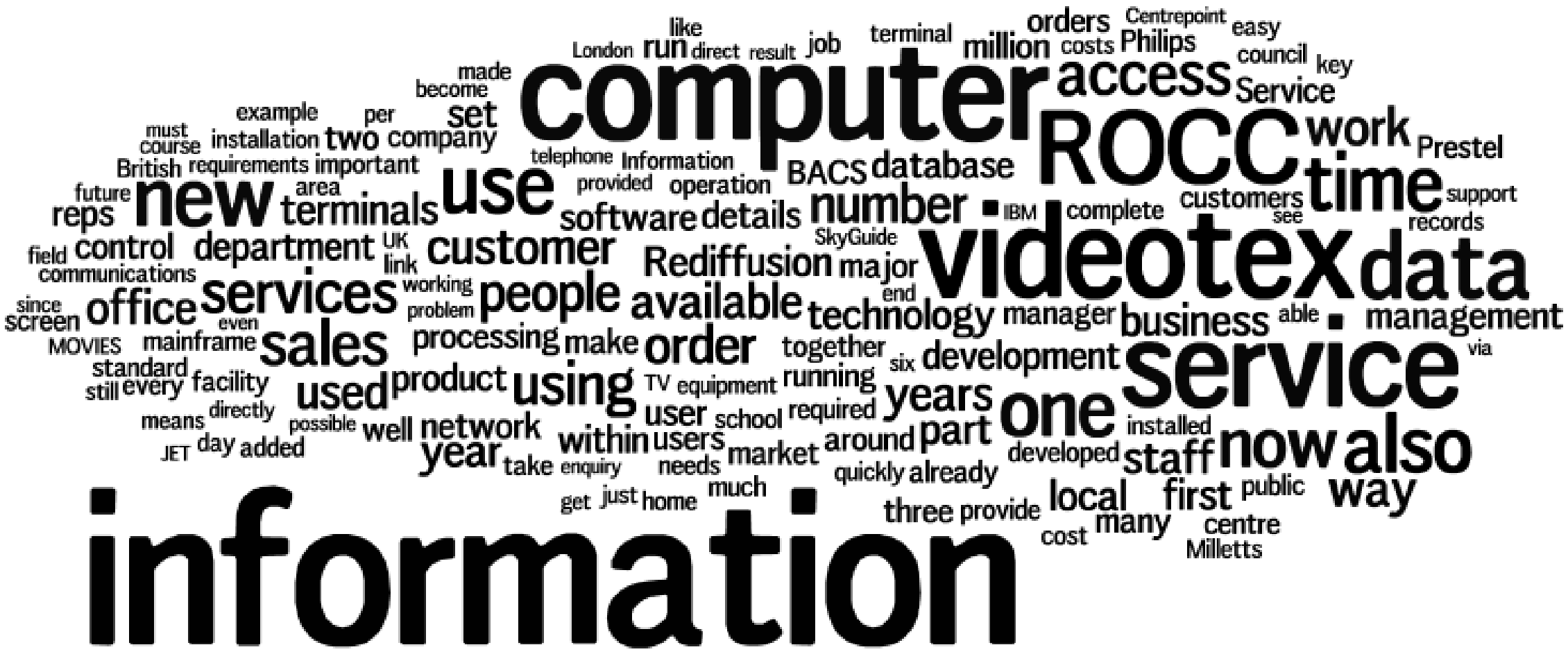

Figure 3: Tag cloud 2 - words ‘system’ and ‘systems’ removed

In the tag cloud below the word ‘information’ has been removed.

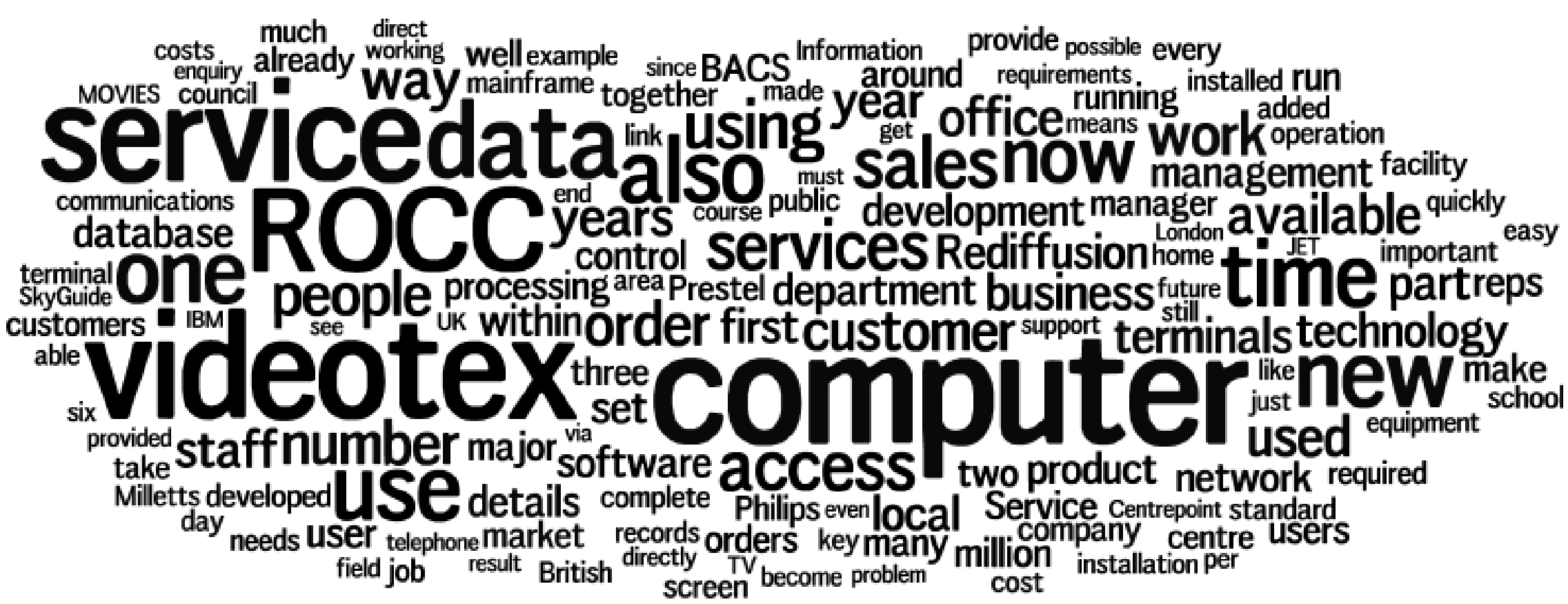

Figure 4: Tag cloud 3 - word ‘information’ removed

**Findings**

There are three main findings in the quantitative analysis. The tag clouds are self-explanatory and especially in tag cloud 1 and 2 the most significant words drown out all other words. After the removal of ‘system’, ‘systems’ and ‘information’, i.e. tag cloud 3, this picture is not that clear anymore. Many other words pop up which make it hard to draw a clear picture of the most important themes.

The first finding is that the case studies focus on the topic information management / information supply of companies. A look at tag cloud 1 shows that the words ‘system’, ‘information’, ‘service’ and ‘data’ are the most significant. Especially the words ‘system’ and ‘information’ are much larger than the other words.

This finding is backed up by tag cloud 2. After the removal of ‘system’, the words ‘information’, ‘data’, ‘time’ and ‘access’ become more significant.

The second finding is the focus on the users of innovative information systems. Words like ‘staff’, ‘user’, ‘terminals’, ‘using’, ‘use’, ‘people’, ‘customers’ and ‘manager’ indicate that. These words are significant in tag cloud 3.

The third finding is that the case studies focus on cost savings with the help of innovative information systems. Words like ‘sales’, ‘cost’, ‘£’, ‘orders’, ‘management’ and ‘services’ indicate that. These words are significant in tag cloud 3 and in the 1-word count.

It is possible, and as mentioned above the aim of this quantitative content analysis, to put these words and findings into different main categories. The three categories which emerged out of this approach are: information management, importance of the user and cost savings.

There is a high emphasis on information management. How high that emphasis is can be highlighted by comparing the three tag clouds. In the first tag cloud the words ‘system’, ‘information’ and ‘service’ stand out unambiguously. After the elimination of the words ‘system’ and ‘systems’ (from tag cloud 2 to 3), the word ‘information’ is the most significant word. But after the elimination of the word ‘information’ (from tag cloud 2 to 3) the relation between the remaining words changes noticeably. In tag cloud 3 many words become more evident and it becomes difficult to identify further categories.

The above mentioned three categories are supported by the 1-word count. Furthermore it needs to be highlighted that several phrase counts were conducted (2-words, 3-words, 4-words and 5-words). These phrase counts are not discussed any further, because they did not deliver further insights. The reason for this is that the phrases are a combination of the words of the 1-word count (and words which can be found in the tag clouds).

### Limitations

Overall the quantitative content analysis has been a useful tool to gather insights into the 29 case studies. On the one side the quantitative analysis does deliver evidence, because the more often a single word is used, the more likely it is to dominate the direction of the marketing material. On the other side the picture which can be drawn from the analysis is rather broad, because it does not show the words in the context they were used. That makes it difficult to interpret the results precisely.

However the limitations of this analysis can be reduced by using a qualitative content analysis - the qualitative analysis is able to check whether the findings can be confirmed or dismissed.

Whether the results of the quantitative content analysis match the results from its qualitative counterpart is an important result. The reason is that any discrepancy or match shows how the case studies market the ‘innovative information systems’. That means it is possible that the case studies focus on one aspect (in this case: information management) by using specific keywords more often than others. But when reading the marketing material the reader gets the feeling that other benefits within the case studies are even more important, they are highlighted by using different keywords or are emphasised indirectly / ‘along the way’.

## QUALITATIVE CONTENT ANALYSIS

### The procedure of the qualitative analysis

The qualitative analysis of innovative information systems used the original case studies in PDF format; it was mainly supported by using the computer software Adobe Acrobat 9 Pro Extended. The software enabled adding comments on the marketing material within the file. No graphic content was lost, because there was no need to convert the PDF files into another format. The software offers several functions, e.g. to highlight relevant text passages and to include sticky notes.

The content analysis was conducted by using a multi-stage design. In the first stage the whole amount (more than 74 pages) was skimmed to get a first impression. In the second stage the whole material was read thoroughly and repeatedly to get an in depth understanding of the case studies. During this stage relevant parts of the text were highlighted and “sticky notes” were included to summarise the findings. This was by far the most time consuming and labour intensive stage of the content analysis.

In the final stage keywords were extracted out of the relevant passages. The passages had to be checked again and again to get an impression of the marketing focus. The keywords and their context were compared with keywords and context found in the quantitative content analysis. Finally a table with keywords could be created.

The rankings in the tables indicate how intensively the keywords have been used. That means the higher a keyword is ranked, the more emphasis the cases put on it. The rankings have not been derived out of a word count. They display the amount, and even more important, the way the keywords have been used, e.g. in headlines, summaries, tables or running text.

**Findings**

*The three different categories*

When conducting the qualitative analysis the same three categories of the quantitative analysis emerged, but with a different emphasis.

The first category deals with direct and indirect cost savings. Direct cost savings means that “save costs” is mentioned directly in the text. An example for this is: ‘Brent Council cuts fleet costs by £200,000, with new Rediffusion-Based Transport Management System’. Indirect cost savings do not mention cost saving directly, but the reader can conclude that costs can be saved by using the IT solution. An example is: ‘We were initially concerned whether videotex, for all its user friendliness, could offer the power we needed to make MINDER a truly useful and timesaving tool’.

The second category emphasises the information supply and information management regarding executives, employees, sales representatives, customers and suppliers. Back in the 1980s it was certainly a major issue to automate processes, save data on hard disk drives and optimise the general information supply within and between companies.

The third category is the focus on the user of the system. The user plays an important rule and this is highlighted regularly. But due to the fact that this marketing material is addressed to board level employees of an organisation the monetary advantages of the IT solutions are the main focus. The advantages for the users are mentioned along the way although the reader can conclude that an easy to use system leads among others to lower costs.

*Category 1 – Cost savings*

The keywords which have been identified in the first category emphasise directly and indirectly cost savings. The following table displays some examples.

Cost savings:

- *AIU, the UK subsidiary of multinational insurance giant American International Group Inc. has made telex a genuinely personal, immediate and simple means of communication while actually REDUCING overall costs by about 15 per cent. They have done it by combing an advanced telex message switching system with their existing in house network of minicomputer screens.*
- *A most important factor which also had to be taken into account was costs - paying for a systems analyst and programmer. Cutler had a budget to adhere to and having paid for his hardware his next big headache was programming costs. Specialists are expensive and usually at the end of the day, more expensive than hardware.*
- *Brent Council cuts fleet costs by £200,000, with new Rediffusion-Based Transport Management System*
- *RNBNY found that the OSCAR system, installed in 1992, significantly reduced costs and increased efficiency. Since that time, however, the bank’s requirements have continued to evolve, while the technology involved in image processing has matured.*

Time savings:

- *"Histograms make the best use of senior management's time in that a glance through a few pages will highlight exceptional results that require further investigation." Where a user accesses a new page, similar in format to the current display - a previous time period or any one of 15 districts for example - only the data changes because the displays share a common mask. This avoids the repainting of the entire screen and considerably reduces the user's waiting time.*

- *Since the time savings and the potential for interactive sales support systems is so significant, Bywater has since identified the ROCC system as capable of opening up a new marketing potential for servicing sales reps of other non-competitive companies needing a speedy order entry and fulfilment distribution system.*

- *"We were initially concerned whether videotex, for all its user friendliness, could offer the power we needed to make MINDER a truly useful and timesaving tool," Wayman said. "In fact, thanks to the efforts of ROCC Computers' software support team at Crawley, we've been able to build in a great deal more power than we thought possible. The hardware is reliable, and the software has definitely delivered the goods."*

*Category 2 – Information management*

The keywords which have been identified in the second category emphasise an optimised information management within organisations. The following table displays some examples.

Information supply, executives & users:

- *In addition to the 'day-to-day' operations there are a number of facilities that are included to provide the statutory statistical information required. This module has been developed in line with the forthcoming changes as recommended by Korner and to allow for speciality costing. Additionally, the LISTER database has an enquiry system that can be used to provide valuable statistical information for use in medical research.*

- *NIIB sells credit to customers mainly through motor dealers, and also directly through bank branches. The company's service quality is maintained through this network by offering a high level of support to dealers, help with training, providing market information, and through the speed of delivery of loans. [...]The detailed costumer information used by Credit-link is additionally helping NIIB acquire knowledge of its customer profiles. This will help the company better target its product marketing now and into the future.*

- *Pupils register - which includes names, addresses and each boy's details. This system is used by the secretarial staff mainly for enquiry purposes. By pressing a few keys, the secretary can in a few seconds, display all the information on 'Freddy Jones'.*

Information supply, customers & public:

- *Using a mix of videotex TVs, Teleputers and VDUs the Telecentre system dealt with all the accreditation procedures for journalists and other specialists reporting on the Papal visit, which resulted in a significant contribution being made to the smooth and efficient administration of the entire operation. The Journalists Accreditation System provided the journalists with a choice of services and Interpress with valuable management information*

- *Journalists and members of the public were given the opportunity of using the videotex TVs and they not only made full use of the 'static' pages of information, covering for e.g. the Pope's itinerary but were able to 'track down' colleagues who were also covering the Papal visit obtaining information on where they were staying. This did not interfere with the work being done by the administration staff.*

- *Currently there are 14 pages of information for the parents to access and this will gradually be extended. Parents with no computer knowledge are 'chuffed' that they have been given a chance to use the school's computer and Bender thinks that this encourages them to influence their children to take an interest in the subject.*

*Category 3 – The importance of the user*

The keyword which has been identified in the third category thematises the importance of the user within organisations. The following table displays some examples.

Easy to use:

- *The volume of data for each period and the need to distribute the information quickly necessitated an electronic solution. Viewdata was the natural choice, since this medium is easy to use and was able to complement the detailed figures with simple histograms.*

- *Andy Wallis says "My team have developed several other techniques to make the EIS both informative and easy to use". For example, the 0 key has been reserved as a 'panic button'. From any point within the EIS route 0 returns the user to the top level menu.*

- *[...] "The R850 had to be a system which would not upset them and which would not change their working methods significantly. Any drastic improvements to their existing system simply to make programming easier would be all very well for me, but Cutler's workforce would have suffered as a result. Consequently, after these in-depth discussions the estimating department now has a first-class, easy to use system, and with some help of their own creation.*

**Summary**

The findings of the qualitative analysis shifted the results of the quantitative analysis from information management towards cost savings. Although the same three categories can be clearly identified, it was noticeable that the category cost savings has higher emphasis than the other two categories.

One of the reasons for this result is where the advantages are mentioned, i.e. running text, headings, graphics and so on. The cost savings are mentioned especially in headings and highlighted text passages. Although not too much is written about them, the reader has the feeling that they are the most important advantage of the innovative information systems. Indirect words like optimised processes and time savings do not state directly that costs can be reduced. But the reader can conclude that these advantages lead to cost savings and in private sector organisations to profit maximisation.

## LOW FREQUENCY AND MISSING WORDS

The discussion above has focussed on the words that appear most often in the cases, but there are also some words and phrases that we might expect to find in the content but do not appear, or are only present with a very low word count.

During the 1980s Professor Michael Porter produced two seminal works, Competitive Strategy (1980) and Competitive Advantage (1985). In 1985 the Harvard Business Review published "How Information Technology Gives You Competitive Advantage" (Porter and Millar) which illustrated the impact that IT could have on the value chain of an organisation. Although the word 'strategy' appears in the cases four times there are no occurrences of 'competitive advantage' or of 'value chain'. So while the works of Porter were having an impact in business schools in the 1980s, that did not transfer immediately into the language used to describe the benefits of information systems.

Other words that might be expected to appear more frequently include innovation (1), productivity (1) and risk (2). The word quality appears 12 times, but never in the phrase 'total quality', despite the focus on Total Quality Management at the time. It is not just that there is an emphasis on tangible financial benefits, discussion of intangibles is completely absent.

It is perhaps less surprising, given that these cases were written before commercial use of the internet, that there are no mentions of 'electronic commerce' or variations such as 'e-commerce' and 'e-business'. Even the word 'internet' does not appear, as the technology most commonly used in the cases (videotex) was capable of being used over public telephone lines.

**CONCLUSION**

Interestingly, the results of the quantitative content analysis indicated that the focus of the marketing material would be on information management / information supply. But the qualitative analysis revealed that the focus is on monetary advantages.

The reason for this difference is that different words and phrases have been used to market the monetary advantages of the IT solutions. The marketing material dealt with case studies, which means that the material itself did not market one solution only - every case is different and unique.

Moreover, there are more possible ways to praise the monetary advantages of the solutions. While information and data are necessary words to describe the advantages of information management / information supply, monetary advantages can be hidden more effectively. To save money, directly and indirectly, can be described with innumerable words / terms: save, pound, £, savings, budget, time, automation, optimise, optimisation, reduce, costs, spending, expenditure, expense, investments, economise, cut down and many more.

Dependent on the type of organisation (public or private sector institution), the utilisation of innovative information systems leads, directly or indirectly, to reduced costs, increased profit or improved service. For public sector organisations the focus is typically on cost savings, because normally they do not aim to maximise profits. For private sector organisations the focus is typically on profit maximisation.

The figure below displays the relations between the different terms. The highlighted boxes display the identified categories. As mentioned before, the box profit maximisation is primarily relevant to private sector organisations.

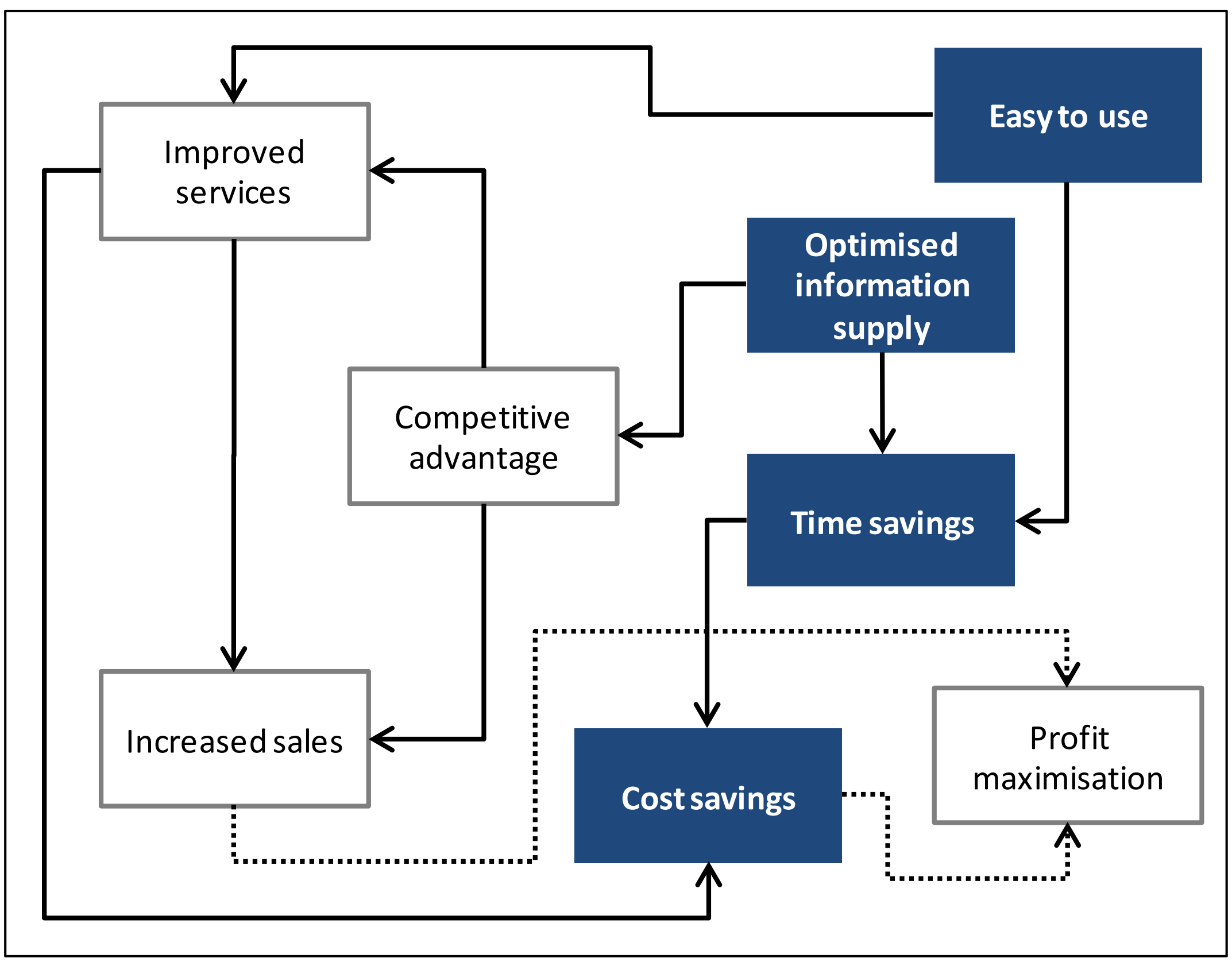

Figure 5: The relation of the keywords identified

In hindsight the strong focus on monetary advantages of information technology seems typical for the 1980s and 1990s. In 1987, Robert Solow stated '[y]ou can see the computer age everywhere but in the productivity statistics' (p. 36). This paradox caused a lot of discussion: 'since the introduction of an 'IT productivity paradox' by Robert Solow (1987), the business value of information technology (IT) has been the topic of many debates by practitioners as well as by academics' (Silvius 2006, p. 93). It took until the 2000s for the IT industry and researchers to realise that there is also a significant amount of intangible value hidden in IT systems (Dedrick & Kraemer 2001; Dedrick, Gurbaxani & Kraemer 2002).

Due to the available amount of time and resources the study had certain limitations. Because of these limitations, there are two recommendations for future research.

Firstly, it would make sense to analyse the marketing material by using a discourse analysis. The difference between the qualitative content analysis and discourse analysis is the level of detail with which data is analysed. When comparing the marketing materials from the 1980s with most recent marketing material on information management, it could show that IT solutions are marketed differently in today's business world. Many of the words and expressions which were used in the cases do not occur in contemporary marketing material. That difference is remarkable, especially to a researcher whose first language is not English.

Secondly, it would make sense to compare the marketing material with recent Business Intelligence marketing material. The reason for this comparison is the similarities of Business Intelligence and innovative information systems 1980 - 1990. Qualitative content analysis of Business Intelligence marketing material (Fass 2013) reveals the potential of such a research endeavour.

Nevertheless, to conduct the double stage content analysis has been an overall success and the results speak for themselves. The disadvantages of both analyses could be minimised and the gap between quantitative and qualitative analysis confirms the value of using both techniques.

**Appendix**

| **Word** | **#** | **Word** | **#** | **Word** | **#** |
|---|---|---|---|---|---|
| | | | | | |
| system | 391 | business | 41 | processing | 32 |
| information | 206 | staff | 41 | cost | 31 |
| service | 141 | terminals | 41 | manager | 30 |
| data | 91 | development | 35 | customers | 26 |
| time | 86 | product | 35 | | |
| systems | 70 | network | 34 | | |
| people | 52 | users | 34 | | |
| work | 49 | £ | 34 | | |
| customer | 47 | market | 34 | | |
| office | 46 | database | 34 | | |
| available | 44 | control | 33 | | |
| technology | 43 | reps | 33 | | |
| management | 42 | link | 32 | | |

Table 1: The 1-word count